\documentclass{article}
\usepackage[utf8]{inputenc}
\usepackage[margin=1in]{geometry}
\usepackage{subcaption}
\usepackage{graphicx}
\usepackage{amsmath}
\usepackage[hyphens]{url}
\usepackage{hyperref}
\hypersetup{colorlinks=false,breaklinks=true,hidelinks}
\usepackage{cleveref}
\usepackage{apacite}
\usepackage{natbib}
\usepackage{authblk}
\usepackage{multirow}
\usepackage{booktabs}
\usepackage{makecell}
\graphicspath{{figures/}}
\title{Seismic Arrival-time Picking on Distributed Acoustic Sensing Data using Semi-supervised Learning}
\author[1]{Weiqiang Zhu\thanks{}, Ettore Biondi, Jiaxuan Li, Jiuxun Yin, Zachary E. Ross, Zhongwen Zhan}
\affil[1]{California Institute of Technology}
\date{}

\begin{document}

\maketitle

\begin{abstract}
Distributed Acoustic Sensing (DAS) is an emerging technology for earthquake monitoring and subsurface imaging. The recorded seismic signals by DAS have several distinct characteristics, such as unknown coupling effects, strong anthropogenic noise, and ultra-dense spatial sampling. These aspects differ from conventional seismic data recorded by seismic networks, making it challenging to utilize DAS at present for seismic monitoring. New data analysis algorithms are needed to extract useful information from DAS data, such as determining the first arrival times of P and S phases for earthquake monitoring and tomography. Previous studies on conventional seismic data demonstrated that deep learning models could achieve performance close to human analysts in picking seismic phases after training on large datasets of manual labels. However, phase picking on DAS data is still a difficult problem due to the lack of manual labels. Further, the differences in mathematical structure between these two data formats, i.e., ultra-dense DAS arrays and sparse seismic networks, make model fine-tuning or transfer learning difficult to implement on DAS data. In this work, we design a new approach using semi-supervised learning to solve the phase-picking task on DAS arrays. We use a pre-trained PhaseNet model as a teacher network to generate noisy labels of P and S arrivals on DAS data and apply the Gaussian mixture model phase association (GaMMA) method to refine these noisy labels to build training datasets. We develop a new deep learning model, PhaseNet-DAS, to process the 2D spatial-temporal data of DAS arrays and train the model using DAS arrays at Long Valley and Ridgecrest, California. The new deep learning model achieves high picking accuracy and good earthquake detection performance. We then apply the model to process continuous data and build earthquake catalogs directly from DAS recording. Our approach using semi-supervised learning provides a way to build effective deep learning models for DAS, which have the potential to improve earthquake monitoring using large-scale fiber networks. 
\end{abstract}

\section*{Introduction}

% broad background
Distributed acoustic sensing (DAS) is a rapidly developing technology that can turn a fiber-optic cable of up to one hundred kilometers into an ultra-dense array of seismic sensors spaced only a few meters apart.
DAS uses an interrogator unit to send laser pulses into an optical fiber and measure the Rayleigh back-scattering from the internal natural flaws of the optical fiber. By measuring the tiny phase changes between repeated pulses, DAS can infer the longitudinal strain or strain rate with time along a fiber-optic cable \citep{zhan2020distributed,lindsey2021fiberoptic,martin2021introduction}.
DAS has proven effective in recording seismic waves for various situations \citep{lindsey2017fiberoptic,williams2019distributed,lindsey2020broadband,li2018pushinga,li2021rapida}.
Compared with traditional forms of seismic acquisition, DAS has several potential advantages in earthquake monitoring. First, DAS can provide unprecedented channel spacing of meters compared with tens-of-kilometers spacing of seismic networks. Second, DAS can take advantage of dark fibers (i.e., unused strands of telecommunication fiber) at a potentially low cost. Third, new DAS interrogator units are becoming capable of longer sensing ranges at a lower cost, as fiber optic networks keep growing with the development of high-speed Internet infrastructure \citep{zhan2020distributed}. 
Thus, DAS is a promising technology for improved earthquake monitoring that is under active research. 
% challenges in data processing of DAS
However, applying DAS to routine earthquake monitoring tasks remains challenging due to the lack of effective algorithms for detecting earthquakes and picking phase arrivals. 
The ultra-high spatial resolution of fiber-optic sensing is a significant advantage compared to seismic networks but also presents a new challenge for traditional data processing algorithms designed for single- or three-component seismometers.
For example, the commonly-used STA/LTA (short-term averaging over long-term averaging) \citep{allen1978automatic} is not effective for DAS because DAS recordings are much noisier than dedicated seismometers due to cable coupling with the ground and strong sensitivity to anthropogenic noise. STA/LTA operates on a single DAS trace and therefore does not effectively use the advantage DAS provides from dense spatial sampling.
Template matching is another effective earthquake detection method, particularly for detecting tiny earthquake signals \citep{gibbons2006detection,peng2009migration,shelly2007nonvolcanic,ross2019searching}. However, the requirement of existing templates and high computational demands limit its applicability for routine earthquake monitoring. 

% deep learning, e.g., phasenet
Deep learning, especially deep neural networks, is currently the state-of-the-art machine learning paradigm for many tasks, such as image classification, object detection, speech recognition, machine translation, text/image generation, and medical image segmentation \citep{lecun2015deep}.
Deep learning is also widely used in earthquake detection  \citep{perol2018convolutionala,ross2018generalized,zhu2019phasenet,mousavi2020earthquakea,zhu2022endtoend,mousavi2022deeplearninga} for studying dense earthquake sequences \citep{park2020machinelearningbasedb,liu2020rapida,tan2021machinea,park2021deepa,su2021high,wilding2022magmatic} and routine monitoring seismicity \citep{huang2020crowdquake,yeck2020leveraging,zhang2022loc,retailleau2022wrapper,shi2022malmi}.
Compared to the STA/LTA method, deep learning is more sensitive to weak signals of small earthquakes and more robust to noisy spikes, which cause false positives for the STA/LTA method.
Compared to the template matching method, deep learning generalizes similarity-based search without the need for precise seismic templates, yet it is also much faster. Neural network models automatically learn to extract common features of earthquake signals from large training datasets and are able to generalize to earthquakes outside the training samples. For example, the PhaseNet model, which is a deep neural network model trained using earthquakes in Northern California, achieves a remarkable performance when applied to tectonic \citep{liu2020rapida,tan2021machinea}, induced \citep{park2020machinelearningbasedb,park2021deepa}, and volcanic earthquakes \citep{retailleau2022automatic,wilding2022magmatica} in multiple places around the world.

% Why deep learning works: training data; or other approaches transfer-learning 
One critical factor in the success of deep learning for earthquake detection and phase picking is the availability of many phase arrival-time measurements manually labeled by human analysts over the past few decades. For example, \citet{ross2018generalized} used $\sim$1.5 million of pairs of P and S picks from the Southern California Seismic Network; \citet{zhu2019phasenet} employed $\sim$700k P and S picks from the Northern California Seismic Network; \citet{michelini2021instance} built a benchmark dataset of $\sim$1.2 million seismic waveforms from the Italian National Seismic Network; \citet{zhao2022diting} formed a benchmark dataset of $\sim$2.3 million seismic waveforms from the China Earthquake Networks; \citet{mousavi2019stanforda} created a global benchmark dataset (STEAD) of $\sim$1.2 million seismic waveforms; Several other benchmark datasets are also created for developing deep learning models \citep{woollam2019convolutional,yeck2020leveraginga,woollam2022seisbench}.
% applications of deep learning to DAS  
Although many DAS datasets have been collected \citep{spica2022pubdasa} and will continue to be collected in the near future, most of these datasets have not yet been analyzed by human analysts. Collecting a large dataset with manual labels for DAS data can be costly and time-consuming.
As a result, there are few applications of deep-learning for DAS data. Most works focus on earthquake detection using a small dataset \citep{hernandez2022improving,lv2022adeneta,huot2022detecting}. Accurately picking phase arrivals is an unsolved challenge for DAS data, hindering its applications to earthquake monitoring.

There have been a number of approaches proposed to train deep learning models with little or no manual labeling, such as data augmentation \citep{zhu2020chapter}, simulating synthetic data \citep{kuang2021realtime,smith2021eikonet,dahmen2022marsquakenet}, fine-tuning and transfer learning \citep{chai2020usinga, jozinovic2022transfer}, self-supervised learning \citep{vandenende2021selfsuperviseda}, and unsupervised learning \citep{mousavi2019unsuperviseda,seydoux2020clustering}.
However, none of those methods have proven effective in picking phase arrival time on DAS data. One reason for this is the difference in the mathematical structure between seismic data and DAS data, i.e., ultra-dense DAS arrays and sparse seismic networks, which makes it difficult to implement model fine-tuning or transfer learning. Additionally, phase arrival-time picking requires high temporal accuracy, which is difficult to achieve through self-supervised or unsupervised learning without accurate manual picks. 
% Semi-supervised learning
Semi-supervised learning is a different approach designed for problems with a small amount of labeled data and a large amount of unlabeled data \citep{xie2020selftraining,zhu2005semisupervised}. There are several ways to use a large amount of unlabeled data as weak supervision to improve model training. One example is the Noisy Student method \citep{xie2020selftraining}, which consists of three main steps: 1) train a teacher model on labeled samples, 2) use the teacher to generate pseudo labels on unlabeled samples, and 3) train a student model on the combination of labeled and pseudo labeled data. Thus, the Noisy Student method can use a large amount of unlabeled data to improve model accuracy and robustness.

% this work
In this work, we present a semi-supervised learning approach for training a deep learning model for seismic arrival-time picking on DAS data without needing manual labels. 
Despite the differences in data modalities between DAS data (i.e., spatio-temporal) and seismic data (i.e., time series), the recorded seismic waveforms exhibit similar characteristics. Based on this connection, we investigate using semi-supervised learning to transfer the knowledge learned by PhaseNet for picking P and S phase arrivals from seismic data to DAS data.
We develop a new neural network model, PhaseNet-DAS, that utilizes spatial and temporal information to leverage hundreds of channels of DAS data to consistently pick seismic arrivals across channels.
We borrow a similar idea of pseudo labeling \citep{arazo2020pseudolabeling} to generate pseudo labels of P and S arrival picks on DAS in order to train deep learning models using unlabeled DAS data.
We extend the semi-supervised learning method to bridge two data modalities of 1D seismic waveforms and 2D DAS recordings so that we can combine the advantages of many manual labels of seismic data and the large volume of DAS data.
We test our method using DAS arrays in Long Valley and Ridgecrest, CA, and evaluate the performance of PhaseNet-DAS in terms of picking precision and recall, phase arrival time resolution, and earthquake detection and location.
% In the following sections, we first introduce the semi-supervised method and the new PhaseNet-DAS architecture, then we evaluate the phase picking performance and present the potential application to earthquake monitoring, and last we discuss the limitations and potential improvements of the current model. 

\section*{Method}

In this section, we discuss the problem of applying deep learning to accurately pick phase arrival times on DAS data in three steps: the semi-supervised learning approach, the PhaseNet-DAS model, and the training dataset.

\subsection*{Semi-supervised Learning}

\begin{figure}
    \centering
    \includegraphics[width=\textwidth]{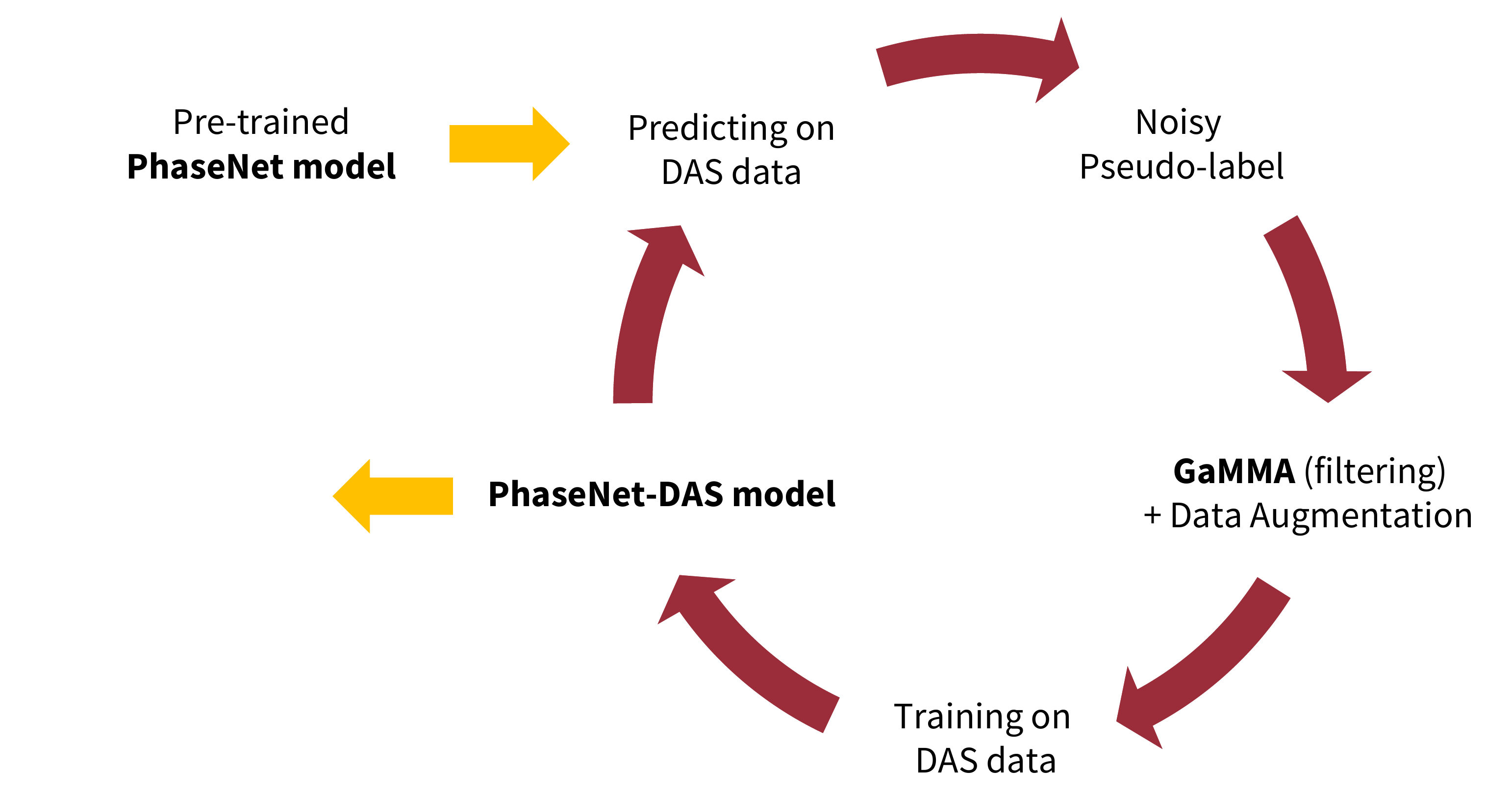}
    \caption{The procedure of semi-supervised learning for training the PhaseNet-DAS model using pseudo-labels generated by the pre-trained PhaseNet model \citep{zhu2019phasenet}. The PhaseNet model is trained using a large dataset of seismic waveforms. This semi-supervised approach can transfer the phase picking capability from PhaseNet to the new PhaseNet-DAS model designed for DAS recordings.}
    \label{fig:semisupervised}
\end{figure}

We explore a semi-supervised learning approach to use unlabeled DAS data to train a deep-learning-based phase picker specifically for DAS. The procedure of the semi-supervised learning approach is summarized in \Cref{fig:semisupervised}.

First, we train a deep-learning-based phase picker on three-component seismic waveforms using many manual picks that have already been labeled by analysts in past decades. Since there are already several widely used deep-learning-based phase pickers \citep{ross2018generalized,zhu2019phasenet,mousavi2020earthquakea}, we directly reuse the pre-trained PhaseNet \citep{zhu2019phasenet} model to omit retraining a deep-learning phase picker for conventional seismic data, which is not the focus of this work.
Although PhaseNet was trained on three-component seismic waveforms, it can also be applied to single-component waveforms because channel dropout (i.e., randomly zero-out one or two channels) is added as a data augmentation \citep{zhu2020seismic}.

Second, we apply the pre-trained PhaseNet model to pick P and S arrivals on each channel of a DAS array independently to generate noisy pseudo labels of P and S picks.
PhaseNet works well on channels with high signal-to-noise (SNR) ratios, but its accuracy is limited compared with the good accuracy on seismic waveforms (\Cref{fig:examples}). For example, the model could detect many false picks from strong anthropogenic noise of DAS data, such as the commonly observed traffic noise. The picked phase arrival times vary significantly between nearby channels since each channel is processed independently. 

Third, we filter the noisy pseudo labels and build a training dataset for DAS data. To accomplish this, we apply the phase association method, Gaussian Mixture Model Associator (GaMMA) \citep{zhu2022earthquake} to filter out false picks from noise and persistent picks across nearby channels. 
GaMMA selects only picks that fall within a small window of the theoretical arrival times of the associated earthquake locations. 
We set the time window size to 1 second. This hyperparameter can be adjusted to balance the trade-off between the quantity and quality of pseudo labels. A small window size results in a small training dataset with high-quality pseudo labels. In contrast, a large window size makes the training dataset large but potentially noisier in arrival times.

Last, we train a new deep-learning-based phase picker designed for DAS data. The model architecture is explained in the following section. 
Because the pseudo labels are mostly picked on high SNR channels, a deep learning picker trained only on high SNR waveforms could generalize poorly to noisy waveforms, which are most common in real DAS data.
Data augmentation, such as superposing noise onto seismic events, can synthesize new training samples with noisy waveforms, significantly expand the training dataset, and improve model generalization on noisy DAS data and weak earthquake signals \citep{shorten2019survey,zhu2020chapter}. Since most DAS data record background noise, we can easily collect many noise samples for data augmentation.
In addition to superposing noise, we added augmentations of randomly flipping data along the spatial axis, masking part of data, superimposing double events, and stretching (resampling) along the time and spatial axis.

By following these steps, we can automatically generate a large dataset of high-quality pseudo labels and train a deep neural network model on DAS data. We can further use this newly trained model to generate new pseudo labels and train an improved model. This procedure can be repeated several times if desired. In this work, we focus on exploring the feasibility of this semi-supervised approach and use only one iteration. The work of optimizing the number of iterations to achieve the best performance can be done in the future.

\subsection*{Neural Network Model}

\begin{figure}
    \centering
    \includegraphics[width=\textwidth]{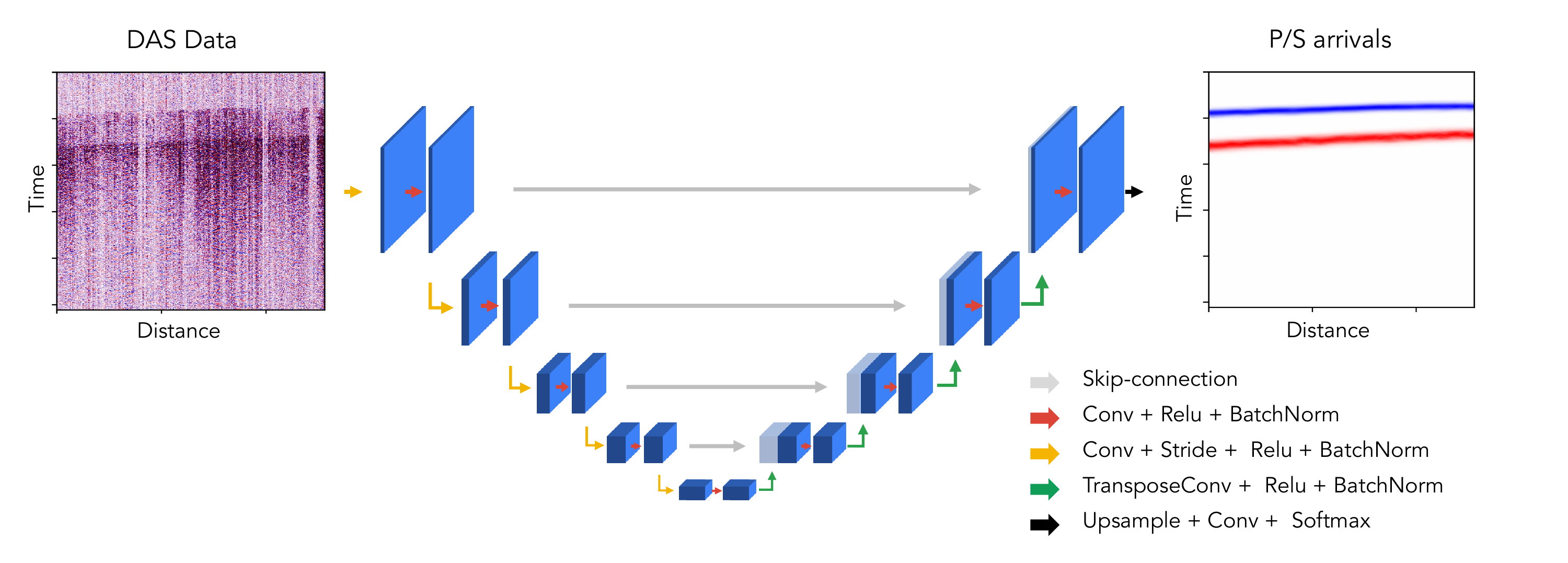}
    \caption{The neural network architecture of PhaseNet-DAS. We use a similar U-Net \citep{ronneberger2015unet} architecture to consider the spatial and temporal information of 2D DAS recordings. PhaseNet-DAS processes raw DAS data through four stages of downsampling and upsampling and a sequence of 2D convolutional layers and relu activation functions, and predicts phase arrivals in each channel of the DAS array as shown by the blue line for P phase and red line for S phase.}
    \label{fig:phasenet_das}
\end{figure}

The pre-trained PhaseNet model is a modified U-Net architecture \citep{ronneberger2015unet} with 1D convolutional layers for processing 1D time series of seismic waveforms. DAS data, on the other hand, are 2D recordings of seismic wavefields with both spatial and temporal information. So the pre-trained PhaseNet model cannot utilize the spatial information from DAS's ultra-dense channels.
In order to exploit both spatial and temporal information of 2D DAS data, we extend the PhaseNet model using 2D convolutional layers. 
The architecture of the new PhaseNet-DAS model is shown in \Cref{fig:phasenet_das}, which is similar to the original U-Net architecture \citep{ronneberger2015unet}. In order to consider the high spatial and temporal resolution of DAS data, we use a larger convolutional kernel size (7$\times$7) and (4$\times$4) stride step to increase the receptive field of PhaseNet-DAS \citep{luo2016understanding}. We add the transposed convolutional layers for up-sampling \citep{noh2015learning}, batch normalization layers \citep{ioffe2015batch}, relu activation functions \citep{glorot2011deep}, and skip connections to the model.
The semi-supervised approach does not require using the same neural network architecture as the pre-trained model, so that we can use other advanced architectures designed for the semantic segmentation task, such as DeepLab \citep{chen2017rethinkinga}, deformable ConvNets \citep{dai2017deformable}, and Swin Transformer \citep{liu2021swinc}.
In this work, we focus on exploring whether we can transfer the knowledge of seismic phase picking from seismic data to DAS data, so we keep a simple U-Net architecture as PhaseNet. The exploration of the best neural network architectures, e.g., transformer \citep{vaswani2017attentiona,mousavi2022deeplearninga,liu2021swinc}, for DAS data can be done in future research.

\subsection*{Training Data}

\begin{figure}
    \centering
    \includegraphics[width=\textwidth]{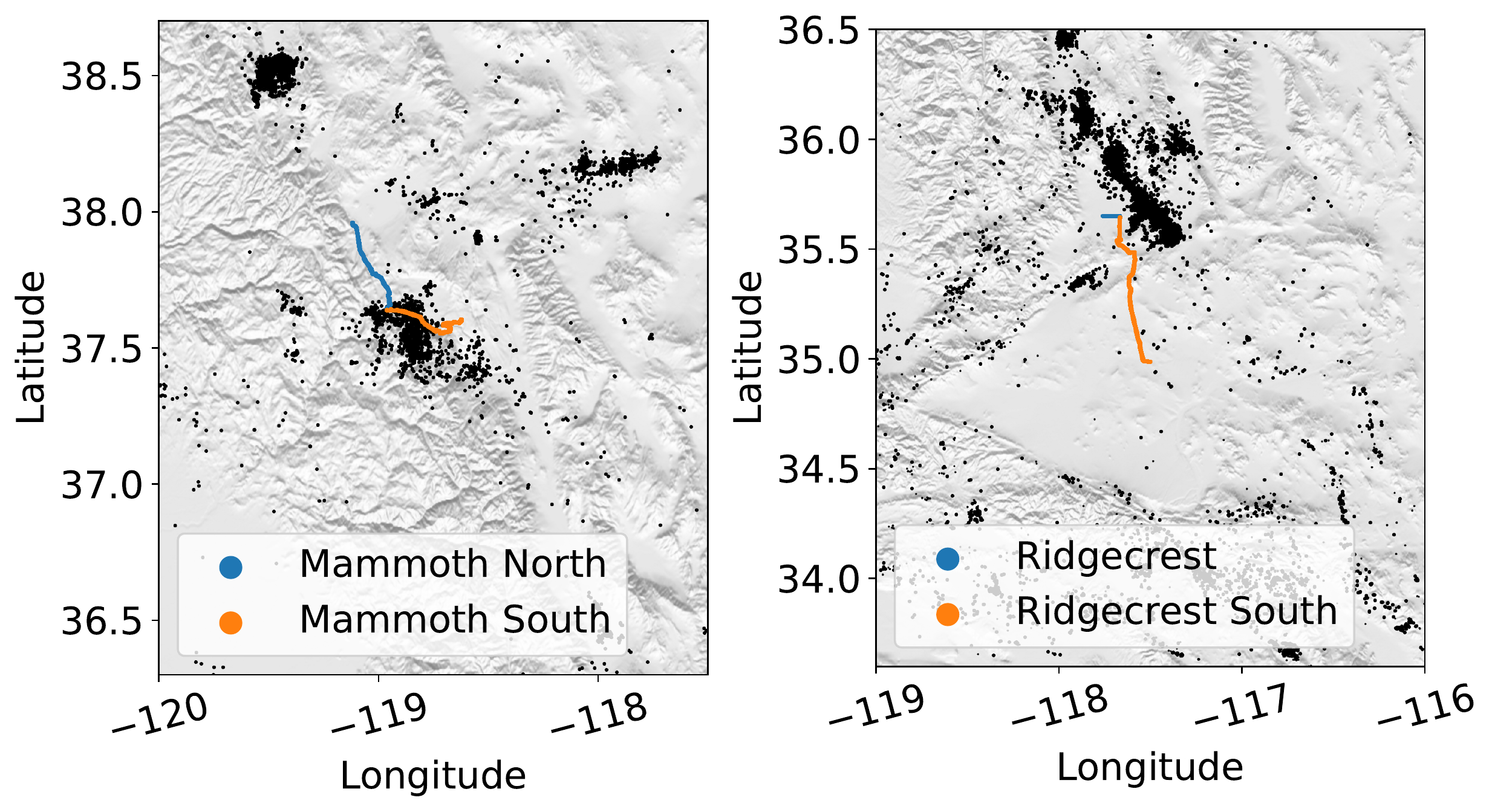}
    \caption{The two DAS arrays used to build the training dataset. The blue and orange lines are the locations of the fiber-optic cables. The black dots are earthquakes in the Northern California earthquake catalog and the Southern California earthquake catalog.}
    \label{fig:das_location}
\end{figure}

We produce a training dataset using two DAS arrays in Long Valley and Ridgecrest, CA (\Cref{fig:das_location}).
The Long Valley DAS array consists of two cables with a total length of 100 km, 10,000 channels, and a spatial resolution of 10 m \citep{li2021rapida,yang2022subkilometera,yang2022faulta,atterholt2022fault}. The Ridgecrest consists of one short cable (10 km and 1,250 channels) and one long cable (100 km and 10,000 channels) \citep{biondi2022geolocalization}.
We retrieved earthquake information from the routine catalogs of the Northern California Seismic Network (NCSN) and Southern California Seismic Network (SCSN) and extracted corresponding DAS records. Following the semi-supervised learning approach outlined previously, we applied the pre-trained PhaseNet model to pick P and S arrivals on these extracted event data, applied the GaMMA model to associate picks, and kept the events with at least 500 P and 500 S picks. Then, we obtain a training dataset of 542 events and 410 events from the north and south Long Valley DAS cables, and 284 events and 419 events from the long and short Ridgecrest cables. 
The corresponding pairs of P and S picks are $\sim$560k and $\sim$352k from the north and south Long Valley DAS cables, and $\sim$182k and $\sim$577k from the long and short Ridgecrest cables. Because we do not have manual labels as ground truth to evaluate the model performance, we only split the dataset into 90\% training and 10\% validation sets. 
We randomly selected a fixed size of 1024$\times$1024 (temporal time $\times$ spatial size) as the input data shape and normalized each channel by removing the mean and dividing by the standard deviation. We train PhaseNet-DAS using the AdamW optimizer and a weight decay of 0.1 \citep{loshchilov2017decoupled}, an initial learning rate of 0.003, a cosine decay learning rate and 100 iterations of linear warm-up \citep{he2019bag}, a batch size of 8, and 3,000 training iterations.

% For a large training datasets, a simple training strategy is commonly used \citep{}.

\section*{Results}

\subsection*{Phase Picking Performance}

\begin{figure}
    \centering
    \begin{subfigure}{0.48\textwidth}
    \includegraphics[width=\textwidth]{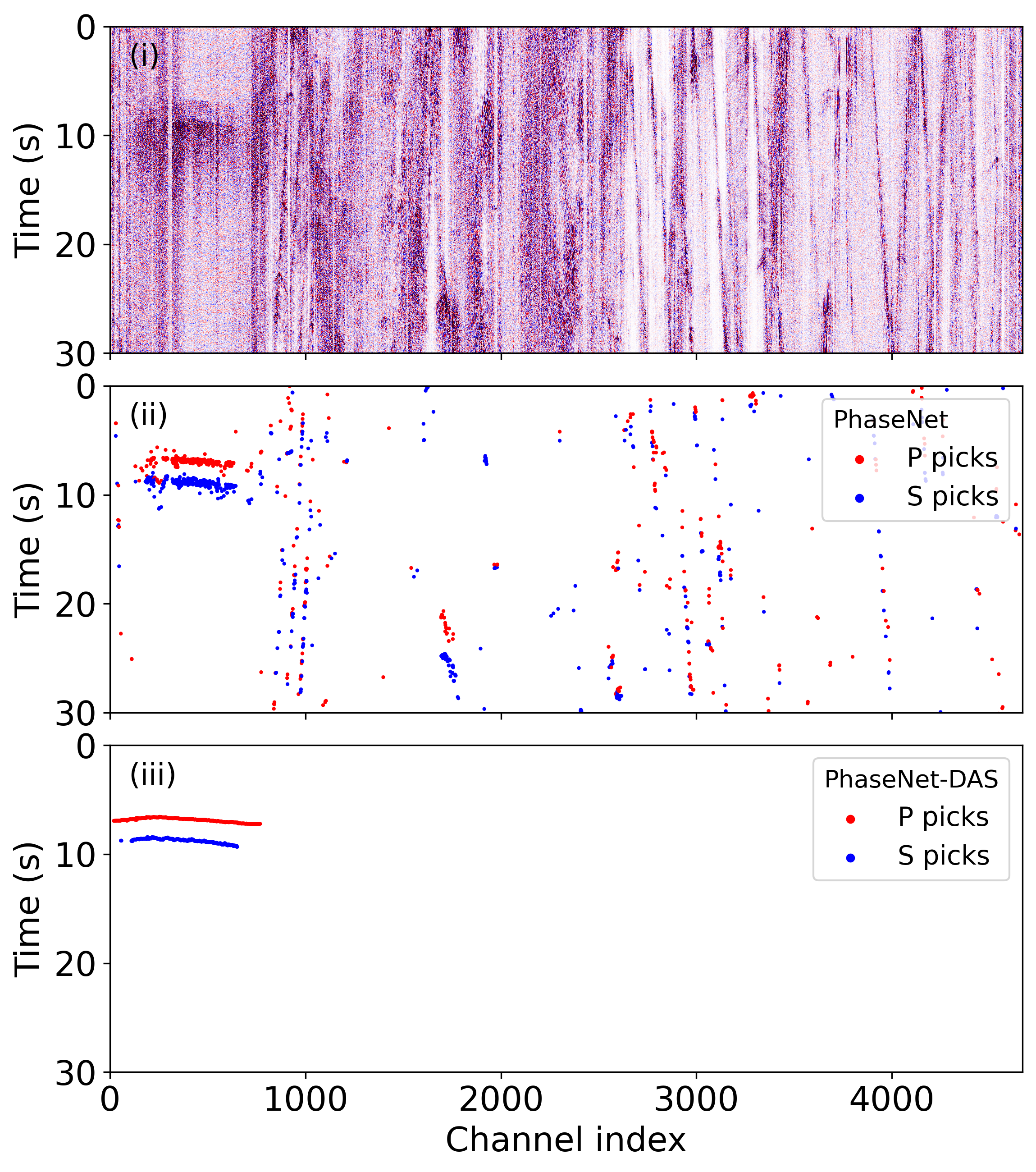}
    \caption{}
    \end{subfigure}
    \begin{subfigure}{0.48\textwidth}
    \includegraphics[width=\textwidth]{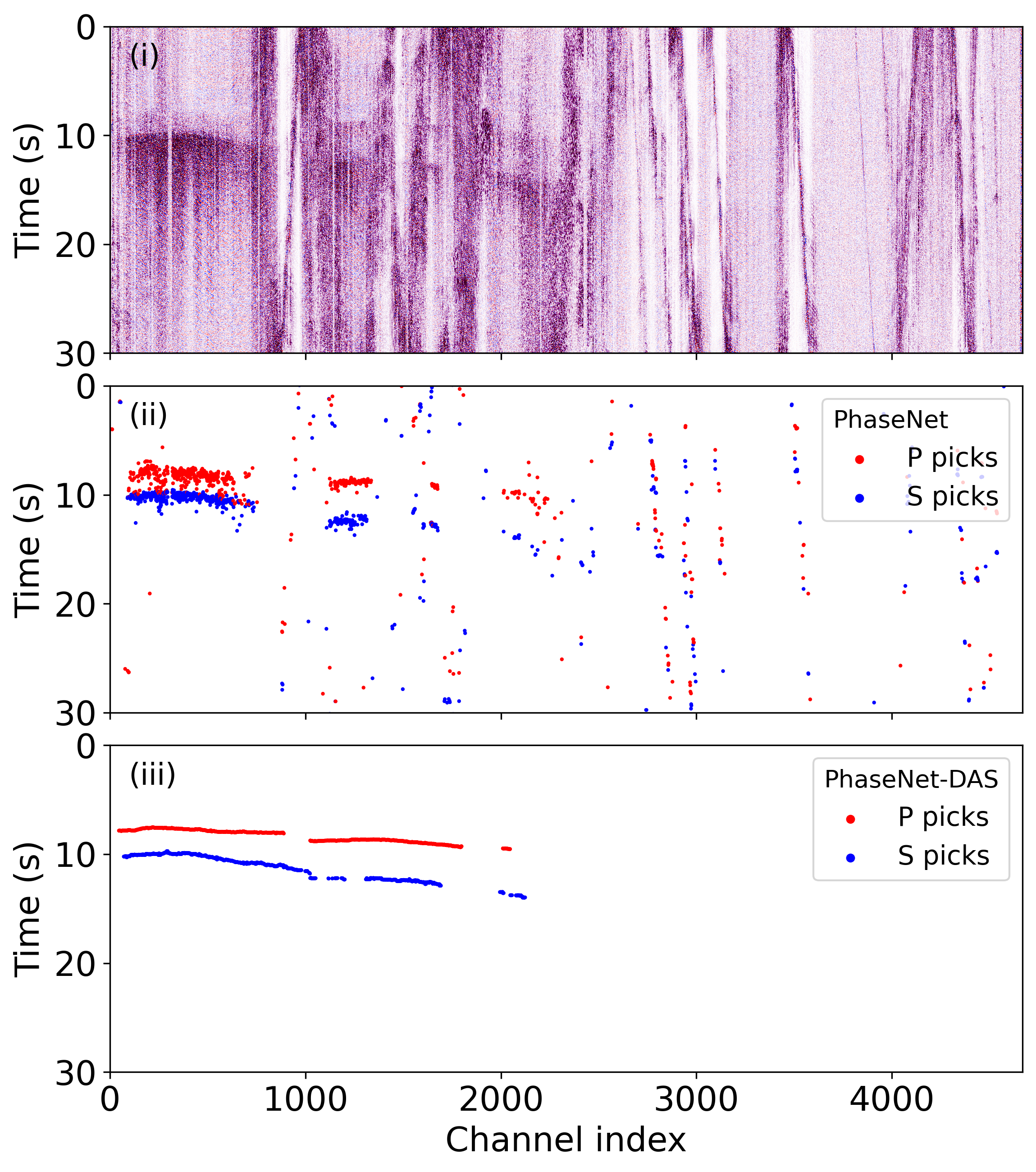}
    \caption{}
    \end{subfigure}
    \begin{subfigure}{0.48\textwidth}
    \includegraphics[width=\textwidth]{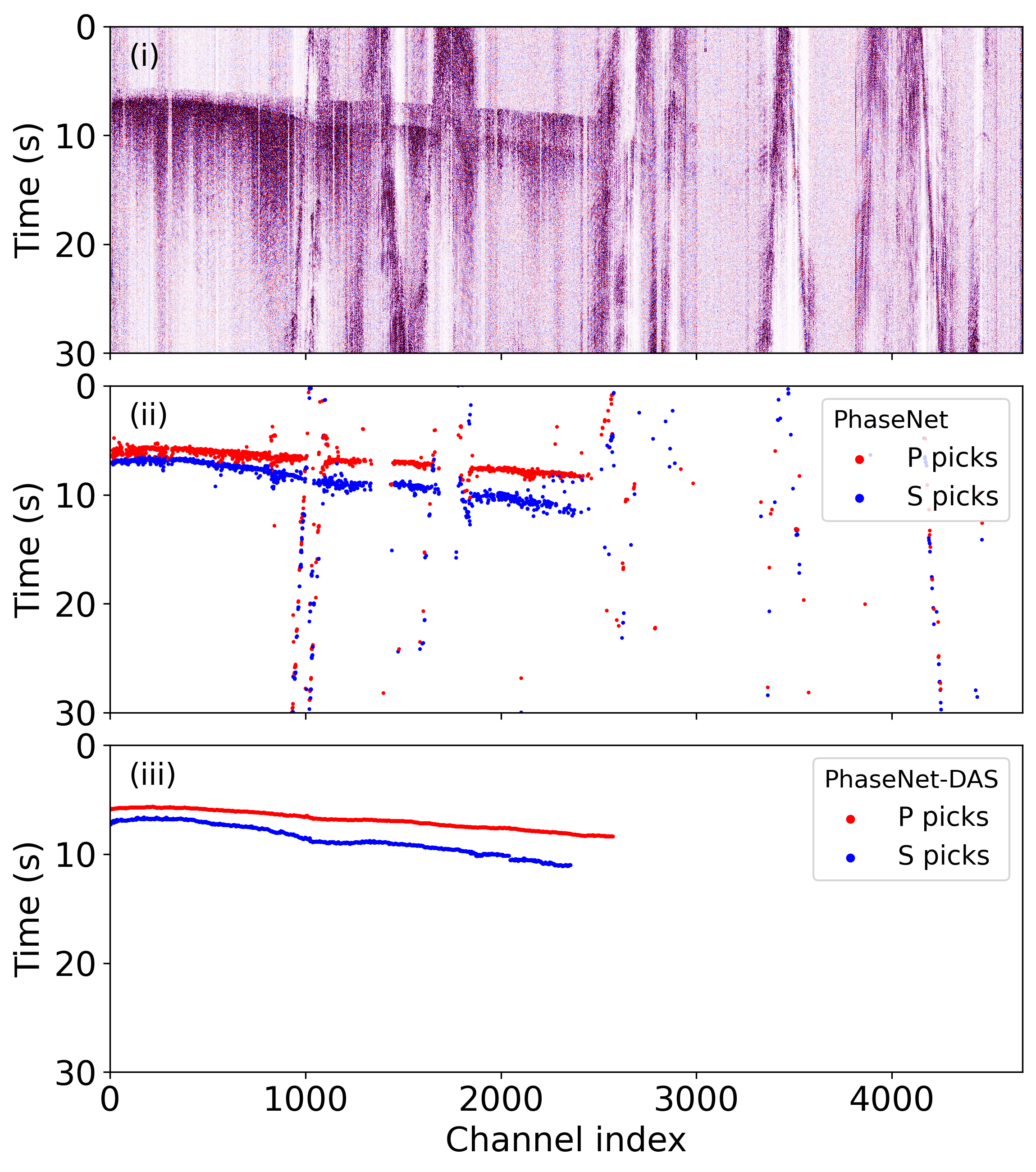}
    \caption{}
    \end{subfigure}
    \begin{subfigure}{0.48\textwidth}
    \includegraphics[width=\textwidth]{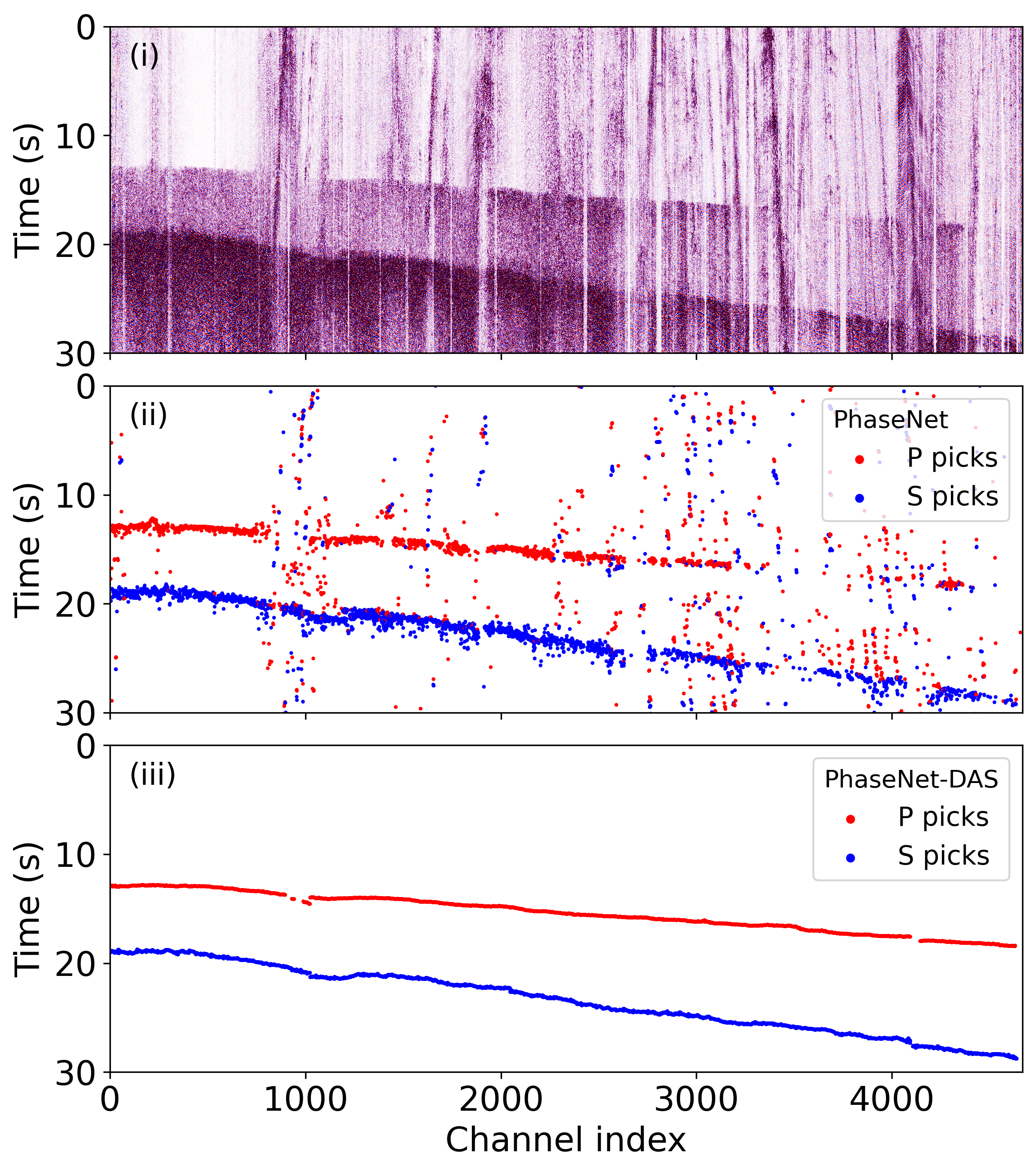}
    \caption{}
    \end{subfigure}
    \caption{Examples of noisy picks predicted by PhaseNet and improved picks prediced by PhaseNet-DAS. Each panel shows (i) DAS recordings of 30 s and 5000 channels, (ii) the PhaseNet picks, and (iii) the PhaseNet-DAS picks.}
    \label{fig:examples}
\end{figure}

% challenges in picking phase arrivals on DAS data. 
%% few false positives
One challenge in picking phase arrivals on DAS data is the presence of strong background noise, as fiber-optic cables are often installed along roads or in urban environments, and DAS is highly sensitive to surface waves. The waveform character of traffic signals has certain resemblance to earthquake signals with sharp emergence of first arrivals and strong surface waves, which leads to many false detections by the pre-trained PhaseNet model. However, when viewed across multiple channels, traffic signals are usually locally visible over short distances of a few kilometers without clear body waves. In contrast, earthquake signals tend to be much stronger and recorded by an entire DAS array with both body and surface seismic waves present. PhaseNet-DAS uses both spatial and temporal information by jointly analyzing multiple channels across a DAS array, making it more robust to traffic noise. \Cref{fig:examples} shows four examples of earthquake signals that can be observed over some parts of the DAS array. Due to strong background noise, we can see that PhaseNet returns many false detections of P and S arrivals. However, PhaseNet-DAS' predictions have fewer incorrect detections and are consistent across channels with less variation in the picked arrival times.

\begin{figure}
    \centering
    % \begin{subfigure}{0.61\textwidth}
    % \includegraphics[width=\textwidth]{mccc_pick_example}
    % \caption{}
    % \end{subfigure}
    \begin{subfigure}{0.71\textwidth}    \includegraphics[width=\textwidth]{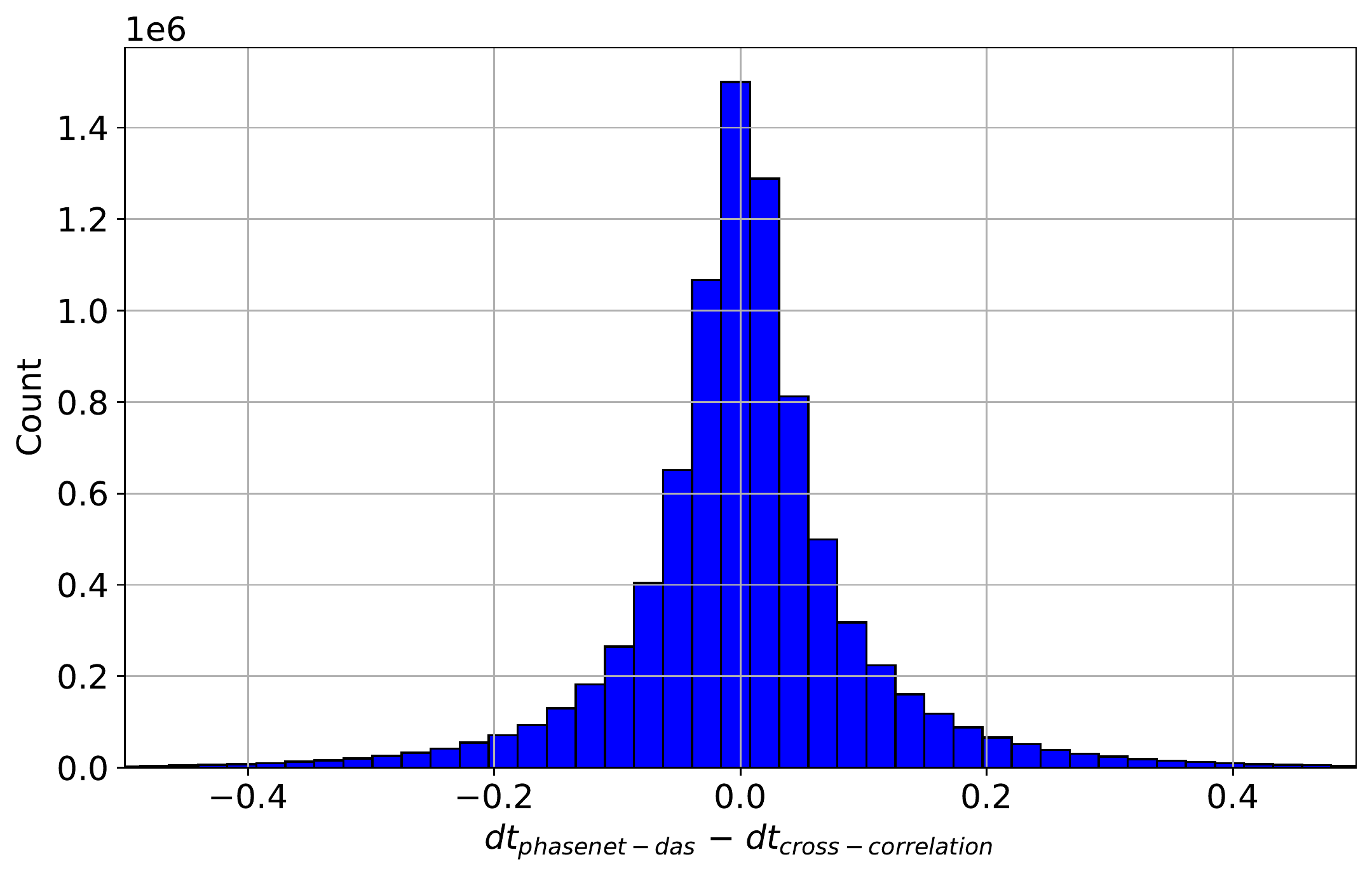}
    \caption{}
    \end{subfigure}
    \begin{subfigure}{0.71\textwidth}
    \includegraphics[width=\textwidth]{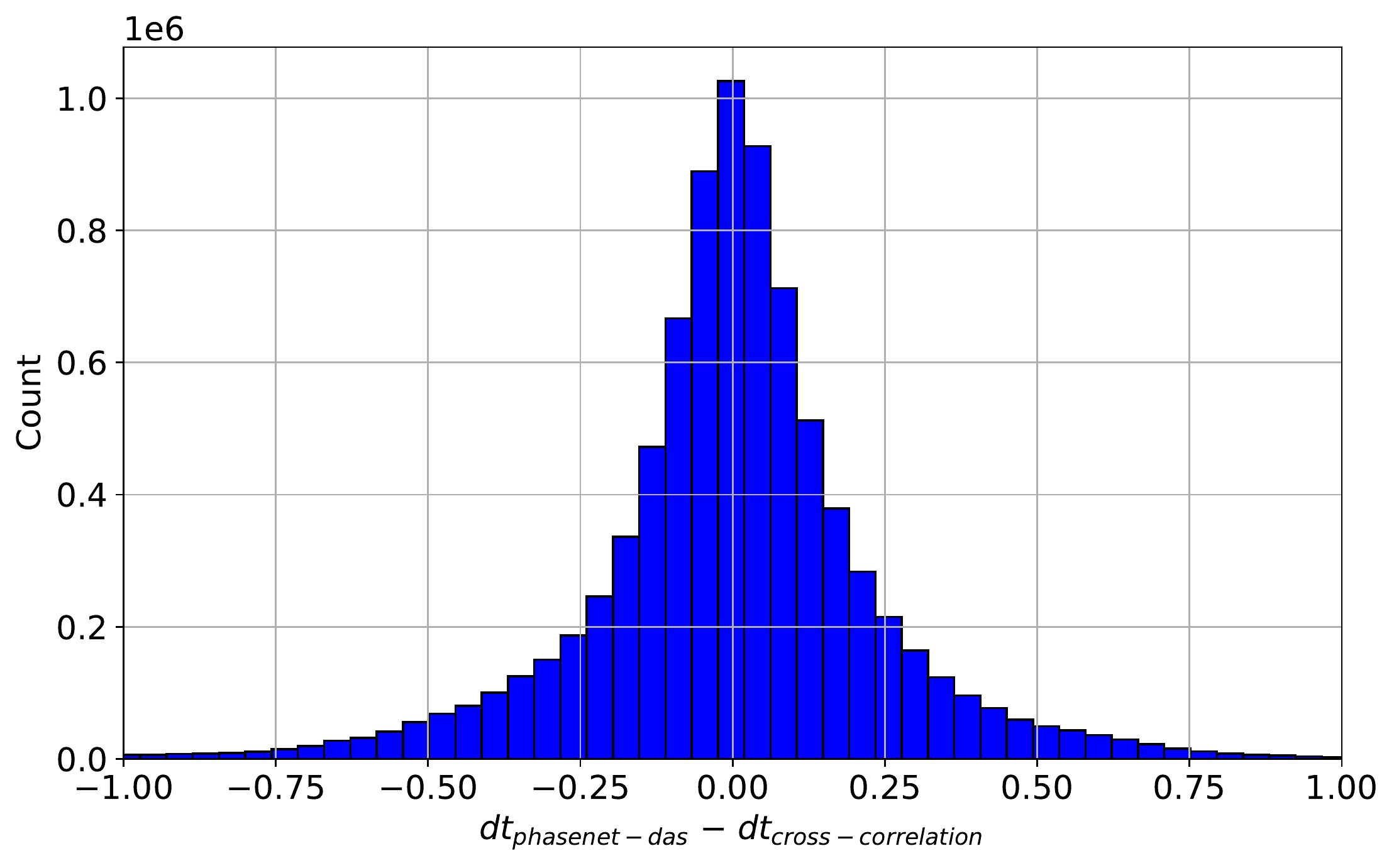}
    \caption{}
    \end{subfigure}
    \caption{Residuals of differential arrival-times picked by PhaseNet-DAS for (a) P waves and (b) S waves. 
    % (a) An example of a representative cross-correlation profile. The green line is the picked residuals of differential arrival-times using multichannel cross-correlation.
    % The residuals of (a) P-wave and (b) S-wave differential arrival times. 
    We first measure differential arrival-times of PhaseNet-DAS picks ($dt_\text{phasenet-das}$) and waveform cross-correlation ($dt_\text{cross-correlation}$) from selected event pairs. Then we can calculate the residuals between these two differential arrival-times ($dt_\text{phasenet-das} - dt_\text{cross-correlation}$) to evaluate the accuracy of PhaseNet-DAS picks, assuming waveform cross-correlation measurement as the ground truth. 
    % (b) The residuals of S-wave differential arrival times.
    }
    \label{fig:differential_time_error}
\end{figure}

%% time accuracy
In addition to traffic noise, other factors such as poor ground coupling and instrumental noise make the signal-noise ratio (SNR) of DAS data generally lower than that of seismic data. The lower DAS SNR hinders the ability to accurately pick phase arrivals with manual labeling or automatic algorithms. Because we do not have manual labels of P and S arrivals, we evaluate the temporal accuracy of PhaseNet-DAS's picks by comparing differential arrival times between two events measured using waveform cross-correlation.
Waveform cross-correlation is commonly used for earthquake detection (known as template matching or match filtering) \citep{gibbons2006detection,peng2009migration,shelly2007nonvolcanic,ross2019searching}, measuring differential travel-time  \citep{waldhauser2000doubledifferencea,zhang2003doubledifference,zhang2015effectivea,trugman2017growclusta}, and measuring relative polarity \citep{shelly2016newa,shellyfracturemesh}. Cross-correlation achieves a high temporal resolution of the waveform sampling rate or super-resolution using interpolation techniques.
We cut a 4s time window around the arrival picked by PhaseNet-DAS, apply a band-pass filter between 1 Hz to 10 Hz, and calculate the cross-correlation between event pairs.
The differential time is determined from the peaks of the cross-correlation profile. Because DAS waveforms are usually much noisier than seismic waveforms and have low cross-correlation coefficients, we further improve the robustness of differential time measurements using multi-channel cross-correlation \citep{vandecar1990determination} to accurately extract the peaks across multiple cross-correlation profiles. We select 2,539 event pairs and $\sim$9 millons differential time measurements for both P and S waves as the reference to evaluate the temporal accuracy of PhaseNet-DAS picks. 
% P wave cross-correlation coefficients higher than 0.37. S wave cross-correlation coefficients higher than 0.6. 
\Cref{fig:differential_time_error} shows the statistics of these two differential time measurements. 
If we assume the differential time measurements by waveform cross-correlation are the ground truth, the errors of differential time measurements by PhaseNet-DAS have a mean of 0.001 s and a standard deviation of 0.06 s for P waves and a mean of 0.005 s and a standard deviation of 0.25 s for S waves.
For comparison, the absolute arrival-time errors of the pre-trained PhaseNet model compared with manual picks have a mean of 0.002 s and a standard deviation of 0.05 s for P waves, and a mean of 0.003 s and a standard deviation of 0.08 s for S waves \citep{zhu2019phasenet}. Although the differential time errors and absolute arrival-time errors can not be directly compared, the similar scales of these errors demonstrate that we can effectively transfer the high picking accuracy of the pre-trained PhaseNet model to the new DAS data.

%% TODO: replot time residual distribution

\subsection*{Applications to Earthquake Monitoring}

\begin{figure}
    \centering
    \includegraphics[width=\textwidth]{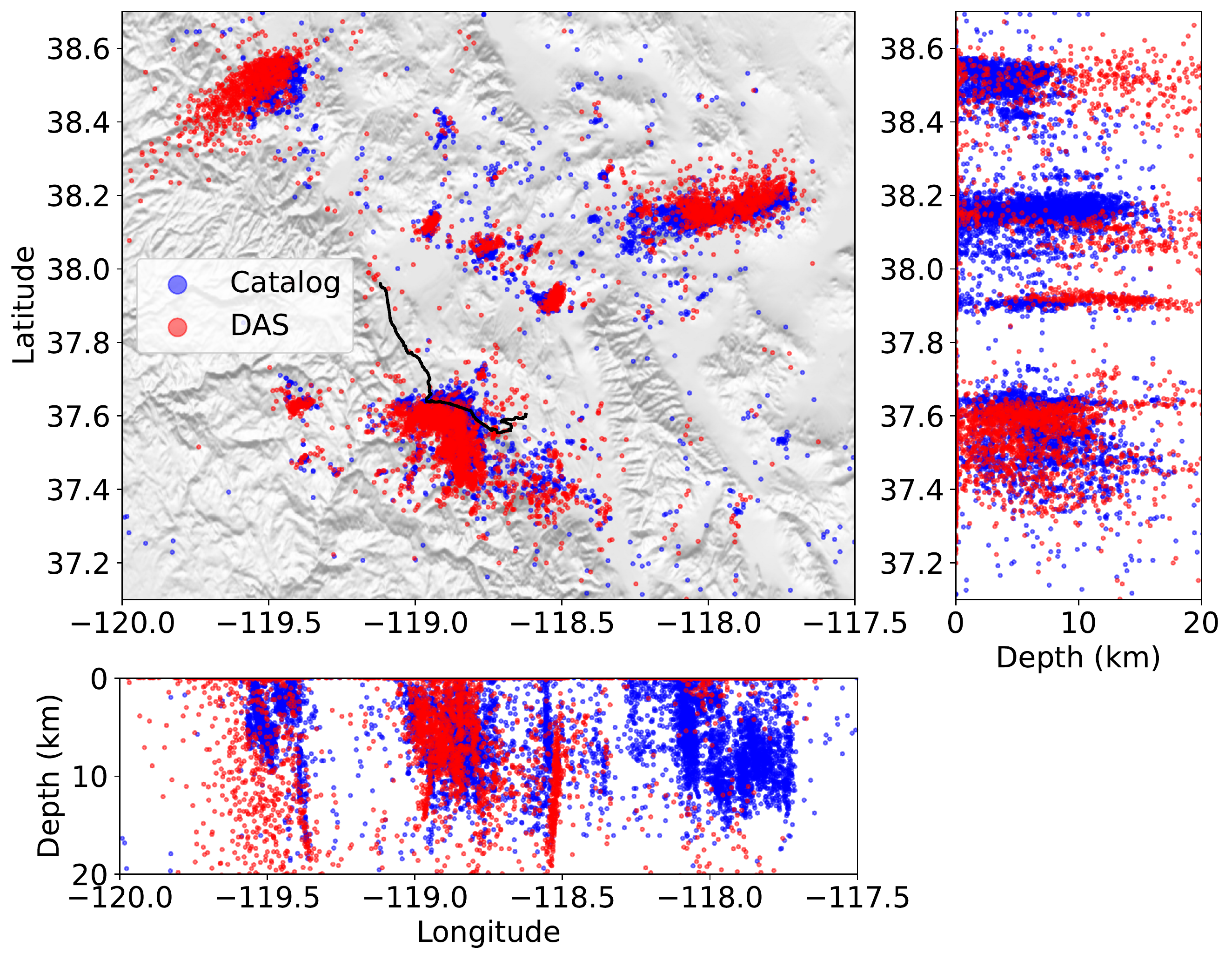}
    \caption{Earthquake locations determined by phase arrival-times measured by PhaseNet-DAS. The blue dots are earthquakes in the Northern California earthquake catalog. The red dots are earthquake determined by the DAS array as shown by the black line.}
    \label{fig:earthquake_location_mammoth}
\end{figure}

\begin{figure}
    \centering
    \includegraphics[width=0.75\textwidth]{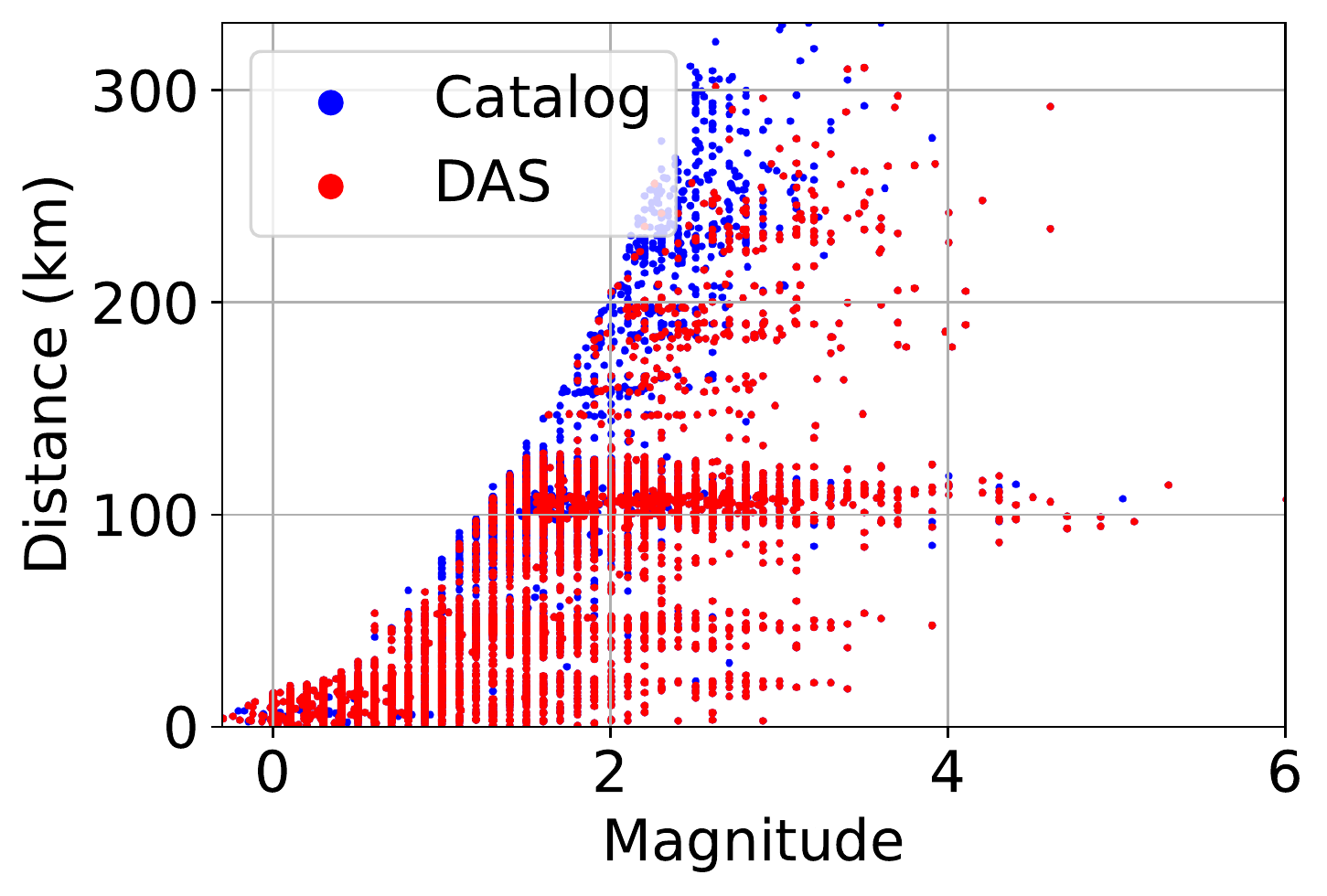}
    \caption{The magnitude distribution of earthquakes detected by PhaseNet-DAS. The blue dots are earthquakes in the Northern California earthquake catalog and the Southern California earthquake catalog within 3 degrees of the Long Valley DAS array. The red dots indicate the earthquakes detected by PhaseNet-DAS with more than 2000 P and S picks.}
    \label{fig:location_recall_mammoth}
\end{figure}

%% demonstrate detection sensitivity
The rapid development of DAS technology and its advantages in high spatial resolution and scalable deployment using existing telecommunication fiber cables make it a promising technology to improve current earthquake monitoring based on seismic networks \citep{zhan2020distributed}. 
One challenge in applying DAS to earthquake monitoring is to develop automatic algorithms to reliably detect earthquakes and accurately measure phase arrival times. 
The experiments above demonstrate that PhaseNet-DAS can effectively detect and pick P- and S-phase arrivals with few false positives and high temporal accuracy, so we can further analyze applying PhaseNet-DAS to earthquake detection. 
Following a similar workflow for seismic datasets \citep{zhu2023quakeflow}, we apply PhaseNet-DAS to DAS records of 9,839 earthquakes in the Northern California earthquake catalog and Southern California earthquake catalog within 3 degrees of the Long Valley DAS array (\Cref{fig:earthquake_location_mammoth}). PhaseNet-DAS detects $\sim$36 million P-picks and $\sim$53 million S-picks from these recordings. Then we apply GaMMA to associate these automatic picks and detect 9,588 earthquakes with more than 2,000 associated P and S picks. Among these events, 65\% events are within 3 s from the cataloged origin times, and 75\% events are within 15 s. 
\Cref{fig:location_recall_mammoth} shows these detected earthquakes' magnitude and distance distribution within 3 s of the cataloged origin times. PhaseNet-DAS can effectively detect small-magnitude events close to the DAS array and most large magnitude ($\ge$ M2) events within 100 km. 
\Cref{fig:earthquake_location_mammoth} shows the approximate locations of these detected earthquakes from event association. The horizontal locations and depth of events within the Long Valley caldera and close to the DAS array can be well-constrained using these automatic arrival time measurements. Due to the limited azimuthal coverage of a single DAS array, hypocenter locations become less constrained with increasing epicentral distance. This physical limitation could be addressed by combining seismic networks, deploying additional DAS arrays, or designing specific fiber geometries in future research.

%% speed
Lastly, we evaluate the prediction speed of PhaseNet-DAS for its potential use in real-time earthquake monitoring and large-scale data mining tasks. DAS is known to be data intensive as a single DAS array can consist of several thousand channels, which poses a challenge for designing efficient algorithms for real-time processing. One advantage of deep learning is the fast prediction speed after training. The rapid development of deep learning frameworks and computing infrastructures such as Graphics Processing Units (GPUs) and Tensor Processing Units (TPUs) makes deep learning a powerful technique for DAS data processing. 
We calculate the prediction speed for a 24-hour DAS recording sampled at 200Hz with 5k channels. The prediction of PhaseNet-DAS only takes 14.7 minutes using 8 GPUs (NVIDIA Quadro RTX 5000). 
Since PhaseNet-DAS is a fully convolutional network (\Cref{fig:phasenet_das}) and the convolution operator is independent of input data size, we can directly apply PhaseNet-DAS to various time lengths and channel numbers depending on the memory limitations of computational servers. Therefore, this prediction task can be embarrassingly parallelized by cutting DAS data into segments of waveforms, and the prediction time can be further reduced using more GPUs.
Meanwhile, the fast prediction speed enables applying PhaseNet-DAS to real-time earthquake monitoring by using sliding windows. Exploring DAS for routine earthquake monitoring or earthquake early warning could be a promising direction for future research.

\section*{Discussion}

% about the method.
DAS technology represents an advancement in observational instruments, providing unprecedented spatial resolution by turning the existing fiber optic infrastructure into dense arrays of sensors to record seismic waveforms.
Meanwhile, deep learning represents an advancement in algorithm development, offering a powerful way to transform historical datasets into effective models for extracting important information from recorded seismic waveforms.
PhaseNet-DAS attempts to combine these two advantages to improve earthquake monitoring and other geophysical applications.
The semi-supervised learning approach bridges the gap between two data modalities, namely conventional seismic traces and DAS recordings, so that we can effectively transfer the phase-picking capability of deep learning models trained on 1D time series of seismic data to the new 2D spatio-temporal measurement of DAS data.

% limitations: bias from PhaseNet can not be corrected.
The experiments above demonstrated the high phase-picking performance of PhaseNet-DAS in consistently detecting phase arrivals across multiple channels of a DAS array and accurately measuring P and S arrival times in each channel. It is also important to note the potential limitations of the current model. While the semi-supervised learning approach addressed the issue of the lack of manual labels for DAS data, the pseudo labels generated by the pre-trained PhaseNet model could potentially be subject to systematic bias, such as missing very weak first arrivals or confusing between phases using single-component data. In order to mitigate these biases, we adopted two approaches in this work. Firstly, we applied phase association to filter out inconsistent phase picks across channels. The phase-picking step using PhaseNet only considers information from a single channel. In contrast, the phase association step incorporates physical constraints across multiple channels, i.e., the phase type should be the same for nearby channels, and the phase arrival time should follow the time move-out determined by channel locations and wave velocities. Through phase association, we reduce the potential bias in pseudo labels of inaccurate phase time or wrong phase types.
Secondly, we added strong data augmentation to the training dataset to increase its size and diversity. For example, we superpose various real noises on the training dataset in order to make the model more sensitive to weak phase arrivals. Because the pseudo labels are generated using data from high SNR events, strong and clear first arrivals are less likely to be missed by PhaseNet. By superposing strong noise, we can make these arrivals similar to the cases of low SNR data with either small magnitude earthquakes or strong background noise, such as during traffic hours. Through such data augmentation, we can reduce the potential bias in the pseudo labels of missing weak arrivals for low SNR events.
Other approaches, such as waveform similarity, could also be considered to further reduce the bias in pseudo labels. Adding regularization, such as Laplacian smoothing between nearby channels to the training loss, could be another direction to reduce the effect of inconsistent labels and improve model performance in future research. 

One common challenge for deep learning is the model generalization to new datasets, as the performance of deep neural network models is closely tied to the training datasets. The current PhaseNet-DAS model was trained using two DAS arrays at Long Valley and Ridgecrest, CA, and the datasets were formatted using a time sampling of 100 Hz and a spatial sampling around 10 m. These factors may limit the model's ability to generalize to other DAS arrays and different data samplings. However, because manual labels of historical seismic data are readily available at many locations, we can also apply the semi-supervised learning approach to train deep learning models for different DAS arrays at different locations.

% broad application: other than earthquake monitoring
Designing effective earthquake detection and phase-picking algorithms is critical for applying DAS to earthquake monitoring, source characterization, subsurface imaging, and other seismic problems. Our work represents a new direction to solve the phase arrival-time picking task on DAS using deep learning. In addition to the examples discussed in this work, the PhaseNet-DAS model can also be used to measure phase arrival times for seismic tomography. The semi-supervised approach could also be applied to developing deep learning models for detecting other seismic signals from DAS data, such as tremors \citep{shelly2007nonvolcanica,beroza2011slow} where large seismic archives are available. 

\section*{Conclusions}
With the deployment of more DAS instruments and the collection of massive DAS datasets, developing novel data processing techniques becomes a key direction in discovering signals and gaining insights from massive DAS data.
Deep learning is widely applied in seismic data processing but has limited applications to DAS data due to the lack of manual labels for training deep neural networks.
We explored a novel approach to applying semi-supervised learning to pick P- and S-phase arrivals on DAS data without manual labels. We applied the pre-trained PhaseNet model to single-component DAS traces to generate noisy phase picks. We further applied the GaMMA model to associate and select consistent picks across multiple traces. We use these picks as pseudo labels for training a new deep neural network model, PhaseNet-DAS, designed for DAS data to pick seismic phase arrivals considering both temporal and spatial information. 
The experiments demonstrate that PhaseNet-DAS can effectively detect P and S arrivals with fewer false picks and similar temporal accuracy compared to the pre-trained PhaseNet model, thus paving the way for applying DAS to earthquake detection, early warning, seismic tomography, and other seismic data analysis. 
The semi-supervised learning approach bridges the gap between the scarcity of training labels for DAS data and the abundance of historical seismic data. This approach enables the development of effective deep learning models for other seismic applications of DAS.

% Table 1: Statistics of the four dataset

% \section*{Data availability}
% The DAS dataset used for training and testing is available from Zhongwen Zhan (zwzhan@caltech.edu) upon request.

% \section*{Code availability}
% The pre-trained model of PhaseNet is available at \url{https://ai4eps.github.io/PhaseNet/}. The model of GaMMA is available at \url{https://ai4eps.github.io/GaMMA/}.
% The source code and pre-trained model of PhaseNet-DAS will be released on Github along with publication. 
% We can share the code and the model with reviewers for testing first.

\section*{Acknowledgements}
We would like to thank James Atterholt for his help in building the training dataset. We would like to thank James Atterholt, Qiushi Zhai, Yan Yang, Jiaqi Fang for their constructive discussions.
We would also like to thank the California Broadband Cooperative for fiber access for the Distributed Acoustic Sensing array used in this experiment. We would like to thank OptaSense for the support provided for this calibration experiment. In particular, the authors thank Martin Karrenbach, Victor Yartsev, and Vlad Bogdanov.
This study is funded by the Gordon Moore Foundation, the National Science Foundation (NSF) through the Faculty Early Career Development (CAREER) award number 1848106, and the United States Geological Survey Earthquake Hazards Program award number G22AP00067.
% Additional funding was provided by the Braun Trust and the United States Geological Survey (USGS) Earthquake Hazards Program (EHP) award number G22AP00067. 

% \section*{Author Constributions}
% W.Z. developed and implemented the algorithm, conducted the experiments and analysis. E.B., Z.R., and Z.Z. co-designed the study. J.L. conducted the picking time error analysis. J.L. and J.Y. built the DAS dataset and tested the model. Z.R. and Z.Z. supervised the project. All authors contributed to writing and editing the manuscript.

% \bibliographystyle{apacite}
\bibliography{references}

\end{document}